\documentstyle[12pt]{article}
\title{On hypergeometric series reductions from integral representations,
the Kamp\'{e} de F\'{e}riet function, and elsewhere}  
\author{Mark W. Coffey\\
Department of Physics\\
Colorado School of Mines\\
Golden, CO  80401\\
(Received $\mbox{~~~~~~~~~~~~~~~~~~~~~~~~~~~~~~~2005}$)}
\date{September 26, 2005}
\begin{document}
\maketitle
\baselineskip=25 pt
\begin{abstract}

Single variable hypergeometric functions $_pF_q$ arise in connection with the power series
solution of the Schr\"{o}dinger equation or in the summation of perturbation expansions
in quantum mechanics.  For these applications, it is of interest to obtain analytic
expressions, and we present the reduction of a number of cases of $_pF_p$ and $_{p+1}F_p$,
mainly for $p=2$ and $p=3$.
These and related series have additional applications in
quantum and statistical physics and chemistry.

\end{abstract}

\vspace{.25cm}
\baselineskip=15pt
\centerline{\bf Key words and phrases}
\medskip

\noindent
hypergeometric series, contiguous relations, reduction formulae, perturbation series 

\vspace{.25cm}
\vfill
\centerline{\bf AMS classification numbers}
33C05, 33C15, 33C20  

\baselineskip=25pt
\pagebreak
\medskip
\centerline{\bf Introduction}
\medskip

The types of single variable hypergeometric series considered here appear in
numerous physical contexts including quantum chemistry, the development of
few- and many-body wavefunctions, and statistical physics 
\cite{gottschalk,niu1,niu2,hussain1}.  A particular instance in statistical
mechanics is a study of phase transitions in the $n$-vector model \cite{hussain1}.
In addition, they have arisen, for instance, in condensed matter physics in the consideration of the effective vortex mass in continuum models of superconductors \cite{coffey}.  Hypergeometric series play an important role in expressing
Clebsch--Gordon and Racah coefficients ($3nj$ symbols) \cite{niu1,niu2}.  We recall
that the orthogonal Hermite and Laguerre polynomials that appear often in standard
quantum mechanical problems are instances of the confluent hypergeometric function $_1F_1$.

Besides these fairly general physical applications, hypergeometric functions of the
form $_qF_q$ or $_{q+1}F_q$ are important in quantum mechanics in the series solution of
the Schr\"{o}dinger equation \cite{gottschalk} and in summing perturbation expansions
\cite{saad}.  We give a few details and offer some remarks that may serve to 
unify earlier treatments.

In Ref. \cite{saad}, Saad and Hall summed perturbation series for the spiked
harmonic oscillator, a linear oscillator with Hamiltonian perturbed by an
inverse power-law term.  In their approach, in order to perform sums with
summands containing either the function $_1F_1$ or $_2F_1$, they needed explicit
expressions for the function $_3F_2$ to insert into an integral representation.
Among their key intermediate results is the expression (\cite{saad}, Lemma 2)
$$z_3F_2(a+1,1,1;2,2;z)={1 \over a} \int_1^{1-z} {{t^a-1} \over {t^a(1-t)}}dt,
~~~~~~a \neq 0, ~~~~~|z| <1.  \eqno(1)$$
We find that if we perform an integration by parts in this equation and then
a change of variable,
$$z_3F_2(a+1,1,1;2,2;z)=-z\int_0^1 {{\ln y} \over {(1-yz)^{a+1}}}dy, \eqno(2)$$
a result that is a special case of a representation found significantly earlier
by Gottschalk and Maslen \cite{gottschalk} in connection with the power series solution
of the Schr\"{o}dinger equation.  In addition, the case of $a=1/2$ in Eq. (2) was
known previously \cite{hussain2}.  

On the other hand, it is still possible to extend some of the integration
results of Gottschalk and Maslen \cite{gottschalk} for the function $_3F_2$ and we
do so in a later section.  These results have a richness of special cases.

In Ref. \cite{exton}, Exton recently investigated reductions of the Kamp\'{e}
de F\'{e}riet double hypergeometric function \cite{appell} and found some new 
relations for single hypergeometric functions.  These reductions arise when the 
two arguments are related by $x = \pm y$ and are essentially the result of the
application of summation formulae for nearly poised hypergeometric series. 
In the following section we describe the particular character of the single variable hypergeometric function special cases of Exton, and point out connections with
earlier work.  We provide direct proofs of these special cases, and the technique
may be useful to other researchers in the variety of physical and engineering
science that we have mentioned.

We then give a more general $_3F_2$ reduction in a succeeding
section.  After completing and illustrating two $_3F_2$ reductions begun by Gottschalk and Maslen \cite{gottschalk}, we supply brief concluding remarks,.

\centerline{\bf Hypergeometric reduction formulae}
\medskip

Exton \cite{exton} deduced the following four relations for single variable hypergeometric functions of the form $_qF_q$ or $_{q+1}F_q$:
$$e^y ~_2F_2(a,1+a/2;a/2,b;-y)=~_2F_2(2+a-b,b-a-1;b,1+a-b;y), \eqno(3)$$
$$(1-y)^{-h} ~_3F_2[h,a,1+a/2;a/2,b;y/(y-1)]=~_3F_2(h,2+a-b,b-a-1;b,1+a-b;y),
\eqno(4)$$
$$_2F_1(a,-e/2;1+a+e/2;x)=(1-x)^e ~_3F_2(a+e,1+a/2+e/2,e/2;1+a+e/2,a/2+e/2;x),
\eqno(5)$$
and
$$_2F_1(a+d-1,-2a;d;y)=(1-y)^{-a} ~_4F_3(a+d-1,-a,2-d-2a,d+2a-1;d,d+a,1-d-2a;y).
\eqno(6)$$
In order for these functions to be well defined, we require that no denominator
parameter be zero or a negative integer.
An examination of Eqs. (3)--(6) indicates their special character.  In each
case, there is one (or more) hypergeometric function(s) with a numerator
parameter exceeding a denominator parameter by precisely one, and this is
exploited in the direct proofs.

Indeed, equations (3)--(6) are partially based upon the identities 
\cite{slater,exton}
$$_3F_2(a,a/2+1,-n;a/2,b;1)={{(2+a-b)_n(b-a-1)_n} \over {(b)_n(a-b+1)_n}},
\eqno(7)$$
$$_3F_2(a,b,-n;a-b+1,2b-n+1;1)={{(a-2b)_n(a/2-b+1)_n(-b)_n} \over {(a-b+1)_n
(a/2-b)_n (-2b)_n}}, \eqno(8)$$
and
$$_4F_3(a,a/2+1,b,-n;a/2,a-b+1,2b-n+1;1)={{(a-2b)_n(-b)_n} \over {(a-b+1)_n
(-2b)_n}}.  \eqno(9)$$
As each of the hypergeometric functions on the left sides of Eqs. (7)--(9)
has a negative integer numerator parameter, they are individually terminating
series at unit argument.

Equations (4) and (5) are proved in Slater's book \cite{slater} (Section 2.4.2)
by means of Vandermonde transformations.  As mentioned there, Eq. (5) is
essentially due to Bailey.

For the direct proofs of Eqs. (3)--(6), we generally transform each side into a
common intermediate expression.  Such alternative intermediate expressions are
likely of independent interest themselves.  As these transformed expressions
have fewer numerator and denominator parameters, they may be suitable for
numerical computations.

Proof of Eq. (3).  For the right side of this equation, we have
$$~_2F_2(2+a-b,b-a-1;b,1+a-b;y)=~_1F_1(b-a-1;b;y)-{y \over b} ~_1F_1(b-a;b+1;y) \eqno(10a)$$
$$=e^y[_1F_1(a+1;b;-y) - {y \over b} ~_1F_1(a+1;b+1;-y),  \eqno(10b)$$
where $_1F_1$ is the confluent hypergeometric function.  This function 
results from a $_2F_2$ function when a numerator parameter equals a denominator
parameter.
In obtaining Eq. (10a), we wrote the defining infinite series for $_2F_2$,
used the relation $(a+1)_j=(1+j/a)(a)_j$, manipulated the series, and applied
the relation $(a)_{j+1}=a(a+1)_j$.  Here, as usual, $(a)_n=\Gamma(a+n)/\Gamma(a)$,
where $\Gamma$ is the Gamma function, denotes the Pochhammer symbol \cite{andrews}.
In obtaining Eq. (10b), we applied Kummer's
first transform $_1F_1(\alpha;\rho;z)=e^z ~_1F_1(\rho-\alpha;\rho;-z)$.  By
again using $(a+1)_j=(1+j/a)(a)_j$, we may rewrite Eqs. (10) as
$$~_2F_2(2+a-b,b-a-1;b,1+a-b;y)=e^y[_1F_1(a;b;-y) - {2 \over b}y
~_1F_1(a+1;b+1;-y)].  \eqno(11)$$

Similarly, by using specifically 
$$(1+a/2)_j = \left [1+{j \over {a/2}} \right ](a/2)_j \eqno(12)$$ 
in the infinite series form of the left side of Eq. (1), we have
$$e^y ~_2F_2(a,1+a/2;a/2,b;-y)=e^y\left[~_1F_1(a;b;-y)-{2 \over b}y
\sum_{j=0}^\infty {{(a+1)_j} \over {(b+1)_j}} {{(-y)^j} \over {j!}}\right]$$
$$=e^y[_1F_1(a;b;-y) - {2 \over b}y ~_1F_1(a+1;b+1;-y)].  \eqno(13)$$
The equality of Eqs. (11) and (13) shows that Eq. (3) holds.

Proof of Eq. (4).  For the right side of this equation, we have
$$_3F_2(h,2+a-b,b-a-1;b,1+a-b;y)=~_2F_1(h,b-a-1;b;y)$$
$$-{h \over b}y ~_2F_1(h+1,b-a;b+1;y).  \eqno(14)$$
By applying a standard transformation of the Gauss hypergeometric function,
$_2F_1(\alpha,\beta;\gamma;z)=(1-z)^{-\alpha}~_2F_1[\alpha,\gamma-\beta;\gamma;
z/(z-1)]$, we have 
$$_3F_2(h,2+a-b,b-a-1;b,1+a-b;y)=(1-y)^{-h}~_2F_1[h,a+1;b;y/(y-1)]$$
$$-{h \over b}y (1-y)^{-(h+1)}~_2F_1[h+1,a+1;b+1;y/(y-1)]$$
$$=(1-y)^{-h}\left[~_2F_1[h,a;b;y/(y-1)]-{{2h} \over b} {y \over {1-y}} ~_2F_1[h+1,a+1;b+1;y/(y-1)]\right ].  \eqno(15)$$
On the other hand, for the left side of Eq. (4), by using Eq. (12) we have
$$(1-y)^{-h} ~_3F_2[h,a,1+a/2;a/2,b;y/(y-1)] = (1-y)^{-h}
\left[~_2F_1[h,a;b;y/(y-1)]\right.$$
$$\left.+{{2h} \over b} {y \over {(y-1)}} ~_2F_1[h+1,a+1;b+1;y/(y-1)] \right ].  \eqno(16)$$
The equality of the expressions in Eqs. (15) and (16) verifies Eq. (4).  We
note explicitly that Exton's Eq. (13) \cite{exton} on the left side has omitted
a $-$ sign in front of the argument $y/(1-y)$.

Proof of Eq. (5).  We proceed similarly to the above proofs, although now we
require contiguous relations \cite{andrews,grad,slater} for the function $_2F_1$.
We first transform the function $_3F_2$ appearing on the right side of Eq. (5):
$$_3F_2(a+e,1+a/2+e/2,e/2;1+a+e/2,a/2+e/2;x) = ~_2F_1(a+e,e/2;a+e/2+1;x)$$
$$+{e \over {a+e/2+1}}x ~_2F_1(a+e+1,e/2+1;a+e/2+2;x).  \eqno(17)$$
We then apply
$$_2F_1(\alpha,\beta;\gamma;z) = (1-z)^{\gamma-\alpha-\beta}  ~_2F_1(\gamma-\alpha,\gamma-\beta;\gamma;z), \eqno(18)$$ 
giving
$$_3F_2(a+e,1+a/2+e/2,e/2;1+a+e/2,a/2+e/2;x) =
(1-x)^{-e}\left[(1-x)~_2F_1(a+1,1-e/2;e/2+1;x)\right.$$
$$\left.+{e \over {a+e/2+1}}x ~_2F_1(a+1,1-e/2;a+e/2+2;x)\right].  \eqno(19)$$ 

We use two different contiguous relations (e.g., \cite{grad}) to re-express each
of the $_2F_1$'s on the right side of Eq. (19).  We have
$$a(1-x)~_2F_1(a+1,1-e/2;a+e/2+1;x)=(a+e/2)~_2F_1(a,-e/2;a+e/2;x)$$
$$-{e \over 2}~_2F_1(a,1-e/2;a+e/2+1;x)$$
$$=(a+e/2)~_2F_1(a,-e/2;a+e/2;x)-{e \over 2} ~_2F_1(a,-e/2;a+e/2+1;x)$$
$$-{{ae} \over {2(a+e/2+1)}} x~_2F_1(a+1,1-e/2;a+e/2+2;x).  \eqno(20)$$
For the second $_2F_1$ on the right side of Eq. (19) we have
$$-{e \over 2}ax ~_2F_1(a+1,1-e/2;a+e/2+2;x)=(a+e/2)(a+e/2+1)\left[~_2F_1(a,-e/2;
a+e/2;x)\right.$$
$$\left.-~_2F_1(a,-e/2;a+e/2+1;x) \right].  \eqno(21)$$ 
This equation is also used to replace the last term of Eq. (20) with $x~_2F_1(...;x)$.
By making all of these substitutions into the right side of Eq. (19) we find
$$_3F_2(a+e,1+a/2+e/2,e/2;1+a+e/2,a/2+e/2;x) = (1-x)^{-e} ~_2F_1(a,-e/2;a+e/2+1;x).
\eqno(22)$$
Therefore, Eq. (5) holds.

Proof of Eq. (6). 
In this case we require an identity for a $_3F_2$ function before reducing to
$_2F_1$ functions.  For the right side of Eq. (6) we have
$$_4F_3(a+d-1,-a,2-d-2a,d+2a-1;d,d+a,1-d-2a;y)=~_3F_2(a+d-1,-a,d+2a-1;d,d+a;y)$$
$$+{{a(a+d-1)} \over {d(a+d)}}y~_3F_2(a+d,1-a,d+2a;d+1,d+a+1;y).  \eqno(23)$$
We have the contiguous relation \cite{rainville,luke}
$$-a(d+2a-1)y~_3F_2(a+d,1-a,d+2a;d+1,d+a+1;y)$$
$$=d(d+a)\left[_3F_2(a+d,-a,d+2a-1;d,d+a;y)-~_3F_2(a+d-1,-a,d+2a-1;d,d+a;y)\right]$$
$$=d(d+a)\left[_2F_1(-a,d+2a-1;d;y) -~_3F_2(a+d-1,-a,d+2a-1;d,d+a;y)\right].  \eqno(24)$$
Then to confirm Eq. (4) we use \cite{luke}
$$_3F_2(a,b,c;d+1,c+1;z)={c \over {c-d}}~_2F_1(a,b;d+1;z)-{d \over {c-d}}~_3F_2(a,b,c;
d,c+1;z),  \eqno(25)$$
and the transformation (18).

\centerline{\bf Other $_3F_2$ function relations}
\medskip

Mostly the above reductions depended upon a Gauss or Clausen hypergeometric function having
a numerator parameter exceeding a denominator parameter by one.  Here we discuss
how this situation can be extended.  The conclusion is that if a numerator
parameter exceeds a denominator parameter by a positive integer $k$, a Clausen $_3F_2$
function may be written as a sum of $k+1$ $_2F_1$ functions \cite{luke}.

As another example to the previous section, we have
$${{(a+2)_n} \over {(a)_n}}=1+2 {n \over a}+{{n(n-1)} \over {a(a+1)}}, \eqno(26)$$
so that
$$_3F_2(b,c,a+2;d,a;x)=\sum_{j=0}^\infty {{(b)_j(c)_j} \over {(d)_j}}\left[1+2{j \over
a}+{{j(j-1)} \over {a(a+1)}}\right ]{x^j \over {j!}}$$
$$=~_2F_1(b,c;d;x)+2{{bc} \over {ad}}x~_2F_1(b+1,c+1;d+1;x)+{{b(b+1)c(c+1)} \over
{a(a+1)d(d+1)}} x^2 ~_2F_1(b+2,c+2;d+2;x).  \eqno(27)$$

In general, we have
$${{(a+k)_n} \over {(a)_n}}={{(a+n)_k} \over {(a)_k}}={1 \over {(a)_k}}\left[(a)_k
+{k \choose 1}n(a+1)_{k-1}+{k \choose 2}n(n-1)(a+2)_{k-2}\right.$$
$$\left. +\ldots+{k \choose k}n(n-1) \cdots (n-k+1)\right], \eqno(28)$$
a result provable by the method of finite differences.  Therefore the general
reduction is given by
$$_3F_2(b,c,a+k;d,a;x)=\sum_{\ell=0}^k {k \choose \ell}{{(b)_\ell (c)_\ell} \over
{(a)_\ell (d)_\ell}} x^\ell ~_2F_1(b+\ell,c+\ell;d+\ell;x).  \eqno(29)$$

\centerline{\bf Other $_3F_2$ function reductions}
\medskip

The authors of Ref. \cite{gottschalk} used identities given in Abramowitz and
Stegun \cite{nbs} to partially transform two particular $_3F_2$ functions.
We complete and then illustrate this development.

We have (\cite{gottschalk}, p. 1987)
$$_3F_2(a,b,b+1/2;a+1,2b;z)={{a4^b} \over z^a}\int_{(1-z)^{1/2}}^1 (1-y)^{a-1}
(1+y)^{a-2b} dy, \eqno(30)$$
and
$$_3F_2(a,b,b+1/2;a+1,1/2;z)={a \over z^a}\int_0^{z^{1/2}} y^{2a-1}[(1+y)^{-2b}
+(1-y)^{-2b}]dy.  \eqno(31)$$
In Eq. (30) or (31), one should take $b \neq 0$ to avoid a trivial case.  In both
Eqs. (30) and (31) one should take $a \neq -1$ to have a well defined $_3F_2$
function.  The form of the integrand factors indicates when certain special
cases may arise:  for instance when $a=1$, $a=2b$, or $b=1/2$ for Eq. (30) and when
$a=1/2$ or $a=1/2-b$ for Eq. (31).  Indeed, we give examples of such cases below. 
It is also obvious that many cases of Eqs. (30) and (31) can be written in terms
of the incomplete Beta function.

We may perform each of the integrations in Eqs. (30) and (31) in terms of
the Gauss hypergeometric function.  We do this by introducing a change of
variable to transform the integral to the interval $[0,1]$, and then apply
a standard integral representation for $_2F_1$ (e.g., 
\cite{nbs,andrews,grad,luke,slater}).  

For Eq. (30), we let $v(y)=[y-(1-z)^{1/2}]/[1-(1-z)^{1/2}]$ and there results
$$_3F_2(a,b,b+1/2;a+1,2b;z)=4^b [1+(1-z)^{1/2}]^{-2b}~_2F_1\left(2b-a,1;a+1;1+{2 \over z} [(1-z)^{1/2}-1]\right).  \eqno(32)$$
We may then use the transformation (e.g., \cite{nbs,grad,luke,slater}) 
$_2F_1(\alpha,\beta;\gamma;z)=(1-z)^{-\alpha}~_2F_1[\alpha,\gamma-\beta;\gamma;
z/(z-1)]$.  After some algebraic manipulations, we obtain the alternative form
$$_3F_2(a,b,b+1/2;a+1,2b;z)={2^a \over {[1+(1-z)^{1/2}]^a}}~_2F_1\left(2b-a,a;a+1;
{1 \over 2}[1-(1-z)^{1/2}]\right ).  \eqno(33)$$
For Eq. (31) we simply use $v(y)=y/z^{1/2}$ and find that
$$_3F_2(a,b,b+1/2;a+1,1/2;z)={1 \over 2}\left[~_2F_1(2b,2a;2a+1;-z^{1/2})+
~_2F_1(2b,2a;2a+1;z^{1/2})\right].  \eqno(34)$$
We have furnished reductions of both Eqs. (30) and (31) for general values of
the parameters $a$ and $b$.

Equations (33) and (34) have a multitude of special cases, several of which return
known results \cite{nbs,grad}.  If $a=1$ in Eq. (33) and we apply a known
reduction for $_2F_1$ \cite{grad}, we obtain
$$_3F_2(1,1-n/2,3/2-n/2;2,2-n;z)={4 \over {nz}}\left[1-2^{-n}(1+(1-z)^{1/2})^n
\right ].  \eqno(35)$$
In this case, $b=1-n/2$ and $n$ need not be an integer.  For the very special
case that $b=1$ in Eq. (33), the reduction $_2F_1(1,1;2;-z)=\ln(1+z)/z$ \cite{grad}
can be applied, giving \cite{hussain2}
$$_3F_2(1,1,3/2;2,2;z)={4 \over z}\left[\ln 2-\ln(1+(1-z)^{1/2})\right]$$
$$=-{4 \over z}\ln\left[{{1+(1-z)^{1/2}} \over 2}\right ].  \eqno(36)$$
When $a=2b$ in Eq. (30) or (33) we have
$$_3F_2(2b,b,b+1/2;2b+1;2b;z)=~_2F_1(b,b+1/2;2b+1;z)={4^b \over z^{2b}}
\left[1-(1-z)^{1/2}\right]^{2b}$$
$$={2^{2b} \over {[1+(1-z)^{1/2}]^{2b}}}.  \eqno(37)$$
This equation is equivalent to a known result \cite{nbs}.

For $a=1/2$ in Eq. (34) we have
$$_3F_2(1/2,b,b+1/2;3/2,1/2;z)=~_2F_1(b,b+1/2;3/2;z)$$
$$={1 \over {2z^{1/2}}}{1 \over {(1-2b)}}\left[(1+z^{1/2})^{1-2b}-(1-z^{1/2})^{1-2b}
\right], \eqno(38)$$
that is also equivalent to a known result \cite{nbs}.  When $b=1$ in this equation,
we have the very special case $_2F_1(1,3/2;3/2;z)=~_1F_0(1;-;z)=1/(1-z)$.
We may inquire as to the case of $b \to 1/2$ in Eq. (38), and in so doing make use of
the expansion in powers of $1-2b$ resulting from writing
$(1 \pm z^{1/2})^{1-2b} = \exp[(1-2b) \ln (1 \pm z^{1/2})]$.  The result is
$$_2F_1(1/2,1;3/2;z)={1 \over {2z^{1/2}}} \ln \left({1+z^{1/2}} \over {1-z^{1/2}}
\right )={1 \over z^{1/2}}\ln \left[{{1+z^{1/2}} \over {(1-z)^{1/2}}}\right], 
~~~~ 0 < z < 1, \eqno(39)$$
equivalent to a previous result obtained in a different way \cite{saad}.
For $b=a+1/2$ in Eq. (34) we obtain
$$_3F_2(a,a+1/2,a+1;a+1,1/2;z)=~_2F_1(a,a+1/2;1/2;z)$$
$$={1 \over 2}\left[_2F_1(2a+1,2a;2a+1;-z^{1/2})+~_2F_1(2a+1,2a;2a+1;z^{1/2})
\right]$$
$$={1 \over 2}\left[_1F_0(2a;-;-z^{1/2})+~_1F_0(2a;-;-z^{1/2})\right ]
={1 \over 2}\left[(1+z^{1/2})^{-2a}+(1-z^{1/2})^{-2a}\right], \eqno(40)$$
a known result \cite{nbs}.  When $b=a-1/2$ in Eqs. (31) and (34) we have
$$_3F_2(a,a-1/2,a;a+1,1/2;z)={1 \over 2}\left[_2F_1(2a-1,2a;2a+1;-z^{1/2})
+~_2F_1(2a-1,2a;2a+1,z^{1/2})\right].  \eqno(41)$$
By the standard transformation (18) (e.g., \cite{nbs,andrews,grad,slater}) the
hypergeometric functions in this equation are given by
$_2F_1(2a-1,2a;2a+1;w)=(1-w)^{2(1-a)}~_2F_1(2,1;2a+1;w)$.
Other cases include $b=1/2-a$ in Eq. (31) and $b=1/2$ in Eq. (30), and still
others may be considered.  Our examples well illustrate the range of
$_3F_2$ reductions obtainable from Eqs. (30)-(34).

\pagebreak
\centerline{\bf Concluding remarks}
\medskip

We have given reductions of some particular single variable hypergeometric
functions.  These and related series appear in a variety of physical science
applications and elsewhere, including the power series solution of the
Schr\"{o}dinger equation and the summation of perturbation expansions in quantum
mechanics.  Our direct proofs of the reductions should serve
to illustrate many of the properties of the functions $_qF_q$ and $_{q+1}F_q$.
More familiarity with such series as we have considered may allow physicists,
quantum chemists, and others to more often use them and to potentially increase
the computational efficiency of numerical methods, for instance, in many-body
wavefunction calculations.

The symbolic implementation of hypergeometric function identities has become
a more and more active area of research \cite{andrews,koepf,koorn}.  When a sum
can be written as a hypergeometric series, many techniques can be brought to
bear, including the W-Z (Wilf-Zeilberger) method \cite{andrews,koepf}.  As 
part of such methods, a recurrence relation is written that shows a "creative
telescoping"' \cite{andrews,koepf}.  Hence new symbolic computing tools 
\cite{koepf,koorn} are becoming available.

\pagebreak

\end{document}